# Piezoelectricity in the dielectric component of nanoscale dielectric/ferroelectric superlattices

Ji Young Jo[1], Rebecca J. Sichel[1], Ho Nyung Lee[2], Serge M. Nakhmanson[3], Eric M. Dufresne[4], and Paul G. Evans[1,*]

[1]*Department of Materials Science and Engineering and Materials Science Program, University of Wisconsin, Madison, Wisconsin 53706, USA*
[2]*Materials Science and Technology Division, Oak Ridge National Laboratory, Oak Ridge, Tennessee 37831, USA*
[3]*Materials Science Division, Argonne National Laboratory, Argonne, Illinois 60439, USA*
[4]*Advanced Photon Source, Argonne National Laboratory, Argonne, Illinois 60439, USA*

The origin of the functional properties of complex oxide superlattices can be resolved using time-resolved synchrotron x-ray diffraction into contributions from the component layers making up the repeating unit. The $CaTiO_3$ layers of a $CaTiO_3/BaTiO_3$ superlattice have a piezoelectric response to an applied electric field, consistent with a large continuous polarization throughout the superlattice. The overall piezoelectric coefficient at large strains, 54 pm/V, agrees with first-principles predictions in which a tetragonal symmetry is imposed on the superlattice by the $SrTiO_3$ substrate.

[*]evans@engr.wisc.edu

Complex oxide superlattices present an opportunity to design structures with nonequilibrium properties that are vastly different from the bulk forms of their components. Superlattices consisting of alternating dielectric and ferroelectric oxides possess an average spontaneous polarization, even with unit-cell-scale thicknesses of the ferroelectric layer, because the electrostatic energy of the structure as a whole is reduced by polarizing the normally unpolarized dielectric [1-4]. The average polarization can exceed the bulk polarization of the ferroelectric component [1], providing a new route for the enhancement of functional properties including piezoelectricity. This average polarization and structural evidence for a static polarization in the non-ferroelectric layers have been observed experimentally [5, 6]. The desirable functional properties of unit-cell-scale superlattices, however, are defined by their responses to applied fields including mechanical stress and electric fields, and have yet to be fully exploited. Fundamentally, the functionality of ferroelectrics arises because electrostatic polarization causes electrical and mechanical phenomena to be strongly coupled [7]. In this Letter, we show that the relationship between polarization and functional properties applies at the nanometer scale in superlattices where there is a large induced polarization in the dielectric component. Our approach allows us to compare the predicted nonequilibrium properties of superlattices with experimental measurements. We find excellent agreement between experiment and first-principles predictions of both structure and piezoelectric response.

In the mean-field free energy description of ferroelectricity the components of the piezoelectric tensor are proportional to the remnant polarization $P$, and to factors quantifying dielectric and electrostrictive properties [7]. The piezoelectric strain, a mechanical response to the applied electric field $E$, provides insight into both electrical and mechanical phenomena. In this sense, the dielectric layers of the superlattice should have a large piezoelectric response arising from their large polarization [8]. Here, we test this prediction by deriving the contributions of the piezoelectric responses of individual components to the overall piezoelectric response of the superlattice using time-resolved synchrotron x-ray microdiffraction as an *in situ* probe of a superlattice capacitor.

The components of the superlattice for our study have well-defined bulk properties: BaTiO$_3$ is a common ferroelectric in its room-temperature tetragonal phase, and CaTiO$_3$ is a centrosymmetric dielectric. The superlattice was prepared by pulsed laser deposition on a 4 nm-thick SrRuO$_3$ bottom electrode on a SrTiO$_3$ substrate [2]. The growth of 480 individual atomic layers, i.e., 80 periods of the 2(BaTiO$_3$)/4(CaTiO$_3$) repeating unit in Fig. 1(a), was monitored using oscillations of the intensity of the specular reflection in reflection high-energy electron diffraction. The in-plane lattice parameter of the superlattice is coherently strained to the SrTiO$_3$ substrate.

The x-ray diffraction pattern of a superlattice, as shown schematically in Fig. 1(b), exhibits a series of reflections with a reciprocal-space separation set by the thickness of the repeating unit. With $E=0$ the average lattice parameter is $t_{avg}$, and superlattice reflections along the specular rod are indexed with $l$ and $m$ such that reflections appear at

$$q_z = \frac{2\pi}{t_{avg}}\left(m + \frac{l}{n}\right),$$ where $m=0,1,2,...$ and $l=...-2, -1, 0, 1, 2,...$. Here $n$ is the total number of atomic layers in the repeating unit and $q_z$ is the reciprocal-space coordinate along the surface-normal direction. For the superlattice shown in Fig. 1(a), the reflection at $l=0$ $m=2$, for which $q_z$ depends only on $t_{avg}$, gives $t_{avg}=3.98$ Å in zero field.

Time-resolved superlattice diffraction patterns were acquired at station 7ID-C of the Advanced Photon Source of Argonne National Laboratory. X rays with a photon energy of 10 keV were focused using a Fresnel zone plate onto a 300 nm spot positioned within a capacitor defined by a 100 μm-diameter Pt top electrode [9]. The electromechanical properties of the superlattice were obtained by applying a triangle-wave electric field $E$. A multichannel scaler synchronized to the applied electric field sorted x-rays detected using an avalanche photodiode detector into 500 counting channels during each electric-field pulse [10]. The process was repeated for several values of $q_z$ to obtain a map of intensity as a function of $q_z$ and time. In order to achieve reasonable counting statistics, the intensity at each $q_z$ was obtained by summing over 20 cycles of the applied electric field.

When $P$ and $E$ are parallel piezoelectric expansion displaces each superlattice reflection to lower $q_z$, as is evident in the diffraction patterns acquired with a peak magnitude of 1.25 MV/cm in Fig. 1(c). Piezoelectricity increases the thickness of the repeating unit by $nt_{avg}\varepsilon$, where $\varepsilon$ is the field-dependent mean strain along the surface-normal direction. The average strain $\varepsilon$ deduced from the $l$=0 $m$=2 reflection is shown in Fig. 2(a). The increase in $\varepsilon$ is proportional to the increase in the electric field for $E$ greater than 0.4 MV/cm, a signature of the overall piezoelectricity of the superlattice. The piezoelectric coefficient $d_{33}$ in the linear regime of the experimentally observed strains in Fig. 2(a) is 54 pm/V. Despite the fact that the bulk dielectric $CaTiO_3$ forms the majority of the superlattice, the value of $d_{33}$ is similar to the piezoelectric coefficients of $BaTiO_3$ films (up to 54 pm/V) [11] and to $Pb(Zr,Ti)O_3$ films (45 pm/V) [12].

We have performed a series of density functional theory (DFT) calculations in which the symmetry of the simulation cell is restricted to the tetragonal space group *P4mm* and the polarization and ionic displacements are constrained to be along [001]. The in-plane lattice parameter was fixed at the calculated lattice parameter of cubic $SrTiO_3$ and the out-of-plane lattice parameter was relaxed to zero stress to simulate the mechanical boundary conditions of the epitaxial superlattice. The calculated polarization [Fig. 2(b)] is nearly constant throughout the repeating unit, with a value of 34 $\mu C/cm^2$ in each $TiO_2$ slab, only slightly less than the 37 $\mu C/cm^2$ predicted for a clamped $BaTiO_3$ film on a $SrTiO_3$ substrate [13]. Calculations predict a piezoelectric response commensurate with this large polarization, yielding a piezoelectric coefficient $d_{33}$ of 51 pm/V.

The closeness of the experimentally observed $d_{33}$, 54 pm/V to the theoretical value of 51 pm/V illustrates an important point. Previously, enhanced piezoelectricity in layered structures has been produced using antiferrodistortive symmetry breaking [14] or at larger length scales using a macroscopic gradient of the polarization across the entire μm-scale thickness of a superlattice [15]. We suspect that in the present case the origin of the large piezoelectric response lies in the symmetry imposed on the system by epitaxial growth. Our calculations find that at low temperatures the *P4mm* structure is unstable with respect to a number of symmetry-lowering structural distortions that could reduce the large polarization in $CaTiO_3$. The agreement between theoretical and experimental values of piezoelectric coefficients here indicates, however, that the tetragonal symmetry of the calculation provides an excellent approximation for the regime in which we measure the piezoelectric response of the superlattice, i.e., applying *E* at room temperature.

The initial nonlinearity of the strain in Fig. 2(a) shows that the highly responsive state is reached only when the system is distorted by a high electric field, above *E*=0.4 MV/cm. This leads to the tantalizing prospect that superlattices can be produced in which electric-field induced phase transitions yield enhanced piezoelectric properties. A second potential origin of the nonlinearity at low fields lies in the decomposition of the polarization of the film into domains at zero field, an effect which has previously been surmised based on the static properties of superlattices [16, 17]. Further investigation will give more detailed insights into understanding the possible role of superlattice phase transitions under applied electric field. These considerations, however, go beyond the scope of this Letter and thus will be discussed elsewhere in the future.

Further indication that the CaTiO$_3$ layers play a crucial role in the piezoelectricity of the superlattice lies in the layer-by-layer origin of the piezoelectric response. The intensities of superlattice reflections result from sampling the structure function of the repeating unit at a small number of points, as in Fig. 3(a). Under nonequilibrium conditions, the structure function, and thus the intensities of superlattice reflections, is changed by the relative displacements of atoms within the repeating unit. This effect provides a route to measure experimentally (i) how the average piezoelectric strain $\varepsilon$ is divided between the two components of the superlattice and (ii) via electromechanical coupling, whether the layer-by-layer polarization is indeed continuous.

In an analytical representation using a sinusoidal modulation of lattice parameters and scattering factors [18], the intensities of the $l=-1$ and $l=+1$ superlattice reflections are proportional to

$$\left[(mn+l)\frac{(t_{BaTiO_3}-t_{CaTiO_3})}{(t_{BaTiO_3}+t_{CaTiO_3})} - l\frac{(f_{BaTiO_3}-f_{CaTiO_3})}{(f_{BaTiO_3}+f_{CaTiO_3})}\right]^2. \qquad (1)$$

Here $t_i$ and $f_i$ are the lattice parameter and the structure factor for component $i$ ($i$=BaTiO$_3$ or CaTiO$_3$). We define $r$ to be the fraction of the average piezoelectric strain arising from distortion in the BaTiO$_3$ component. When $r = n_{BaTiO_3} t_{BaTiO_3}(\varepsilon=0)/nt_{avg} \approx n_{BaTiO_3}/n = 1/3$ with $t_{BaTiO_3}(\varepsilon=0)/t_{avg}$ close to 1, both components have equal strain. In terms of $r$, the lattice parameters in the BaTiO$_3$ and CaTiO$_3$ layers are $t_{BaTiO_3}(\varepsilon) = t_{BaTiO_3}(\varepsilon=0) + r\varepsilon t_{avg} n/n_{BaTiO_3}$ and $t_{CaTiO_3}(\varepsilon) = t_{CaTiO_3}(\varepsilon=0) + (1-r)\varepsilon t_{avg} n/n_{CaTiO_3}$. For $\varepsilon \ll 1$, the change in intensity with increasing $\varepsilon$ includes only terms proportional to

$$\varepsilon\left[r - \frac{n_{BaTiO_3}}{n}\right], \qquad (2)$$

and to the square of this quantity. Expression (2) predicts that for $r=1/3$, when $BaTiO_3$ and $CaTiO_3$ are equally strained, the intensities of the $l=-1$ and the $l=+1$ satellite reflections will not be changed by piezoelectric expansion.

Profiles of the $l=-1$ and $l=+1$ reflections at $m=2$ are shown in Figs. 3(b) and (c), for zero field and $E=1$ MV/cm, respectively. Both reflections decrease in peak intensity and broaden at high $E$. We attribute the broadening to inhomogeneity of the piezoelectric response or electric field either within the lateral spot size of the focused x-ray beam or across the thickness of the superlattice. At $E=1$ MV/cm, corresponding to an average strain of 0.45%, the change in the integrated intensity with respect to zero field is +2% for the $l=-1$ satellite and -4% for the $l=+1$ satellite.

A numerical kinematic diffraction calculation provides the intensities of the $l=-1$ and $l=+1$ reflections as a continuous function of $r$. This model differs from the sinusoidal approximation in that it uses atomic positions derived from the zero-field DFT calculations and extrapolates to nonzero $E$ by stretching the $CaTiO_3$ and $BaTiO_3$ components according to the parameter $r$. Simulated and experimentally observed changes in the intensities of the $l=-1$ and $l=+1$ reflections are shown as a function of $\varepsilon$ and $r$ in Figs. 4(a) and (b). The best agreement between the kinematic diffraction calculations and the observed small changes in intensity occurs when $r$ is close to 1/3, at which the dielectric and ferroelectric layers have equal piezoelectric response.

A hypothetical superlattice composed of materials with their bulk properties, in which $r=1$, and only the $BaTiO_3$ has spontaneous polarization and resulting piezoelectricity is an extremely poor fit for the experimental results. In the hypothetical case, kinematic diffraction predicts that the $l=-1$ and the $l=+1$ reflections at $m=2$ would increase in intensity by 20% and 120%, respectively at $\varepsilon=0.45\%$, an effect clearly not present in the data. The opposite limiting case of $r=0$, corresponding to localizing the piezoelectric strain in $CaTiO_3$, would lead to a decrease in intensity by 45% at $l$=+1, which is similarly not observed. A large fraction of the piezoelectric distortion thus unambiguously occurs in the CaTiO$_3$ component of the superlattice.

Equal magnitudes of the piezoelectric strains in BaTiO$_3$ and CaTiO$_3$ are consistent with mean-field expectation that large, nearly equal, polarizations in these layers produce piezoelectric coefficients of similar magnitudes in a simple assumption on the continuity of permittivity and electrostrictive coefficient. This prediction of the functional properties using the continuity of the polarization can be compared with the atomic-scale structure derived from DFT calculations. Figure 4(c) compares DFT calculations of the fractional changes of Ca-Ca and Ba-Ba distances for average strains of 0.5% and 1% with the changes expected from a uniform distortion with $r$=1/3. The uniform division of the distortion between BaTiO$_3$ and CaTiO$_3$ is in excellent agreement with the DFT results for strains up to $\varepsilon$=0.5%, which is consistent with the experimental results.

An approach similar to the one we have demonstrated here, combining DFT calculations and nonequilibrium structural probes, can be used to probe the stabilization of structural phases in low dimensional systems [19], the coupling between ferroelectricity and magnetism via structural distortions [20], and to develop new methods to control thermal properties in dielectrics using compositional grading [21]. Our results show that even in these symmetric superlattices there can be an important role of the imposed crystallographic symmetry in determining the piezoelectric response. The role of compositional symmetry breaking in the properties of superlattices can likewise be resolved by a similar approach [22]. The functionality of complex oxides is now being engineered with layer thicknesses at which these new approaches are necessary to extend the conventional volume-average characterization of the properties of these materials to far smaller spatial scales.

This work was supported by the U.S. Department of Energy, Office of Basic Energy Sciences, through Contract No. DE-FG02-04ER46147. H.N.L. acknowledges support from the Division of Materials Sciences and Engineering, U.S. Department of Energy, through Contract No. DE-AC05-00OR22725. S.M.N. and the use of the Advanced Photon Source were supported by the U. S. Department of Energy, Office of Science, Office of Basic Energy Sciences, under Contract No. DE-AC02-06CH11357.

**Figure Legends**

**Figure 1.** (Color online) (a) One repeating unit of a superlattice consisting of 2 unit cells of BaTiO$_3$ and 4 unit cells of CaTiO$_3$. (b) Schematic x-ray diffraction pattern with superlattice reflections along the $q_z$ axis of reciprocal space for the steady state structure at $E$=0 and for the piezoelectrically distorted superlattice at nonzero $E$. (c) Time-dependent x-ray diffraction patterns for the $l$=-3, -2, -1, 0, and 1 reflections of the superlattice with $m$=2, and the (002) reflection of the SrTiO$_3$ substrate. The time dependence of the applied electric field $E$ appears in the leftmost panel. The (002) reflection of the SrTiO$_3$ substrate is unchanged by the applied electric field.

**Figure 2.** (Color online) (a) Average piezoelectric strain $\varepsilon$ as a function of applied electric field $E$, observed using the shift in $q_z$ of the superlattice reflection at $l$=0 $m$=2. (b) First-principles calculation of the layer-by-layer polarization $P$ in the TiO$_2$ slabs within a single repeating unit of the superlattice. The polarization calculated for a strained BaTiO$_3$ with an in-plane lattice parameter matching a SrTiO$_3$ substrate (thin blue line) is only slightly larger than that of the BaTiO$_3$/CaTiO$_3$ superlattice [13].

**Figure 3.** (Color online) (a) The intensity of superlattice reflections (red solid line) is determined by sampling the structure function of the repeating unit (black dashed line) at integer values of $l$. Electric-field dependence of the (b) $l$=-1 $m$=2 and (c) $l$=+1 $m$=2 reflections. The shaded regions represent the area integrated to obtain the integrated intensity.

**Figure 4.** (Color online) Kinematic simulation of the change in the intensities of the (a) $l$=-1 $m$=2 and (b) $l$=+1 $m$=2 reflections, as a function of the average strain $\varepsilon$ and the fraction $r$ of the strain occurring in BaTiO$_3$. Symbols represent the intersection of the experimental strain and integrated intensity with the simulated intensity. The solid lines indicate the limits set by the statistical uncertainty in the integrated intensity. The experimental results are consistent

with $r$=1/3, corresponding to equal strains in the BaTiO$_3$ and CaTiO$_3$ components. (c) DFT predictions of the fractional change in the Ca-Ca and Ba-Ba distances (symbols) are consistent with the sharing of strain according to $r$=1/3.

Jo *et al.*, Figure 1

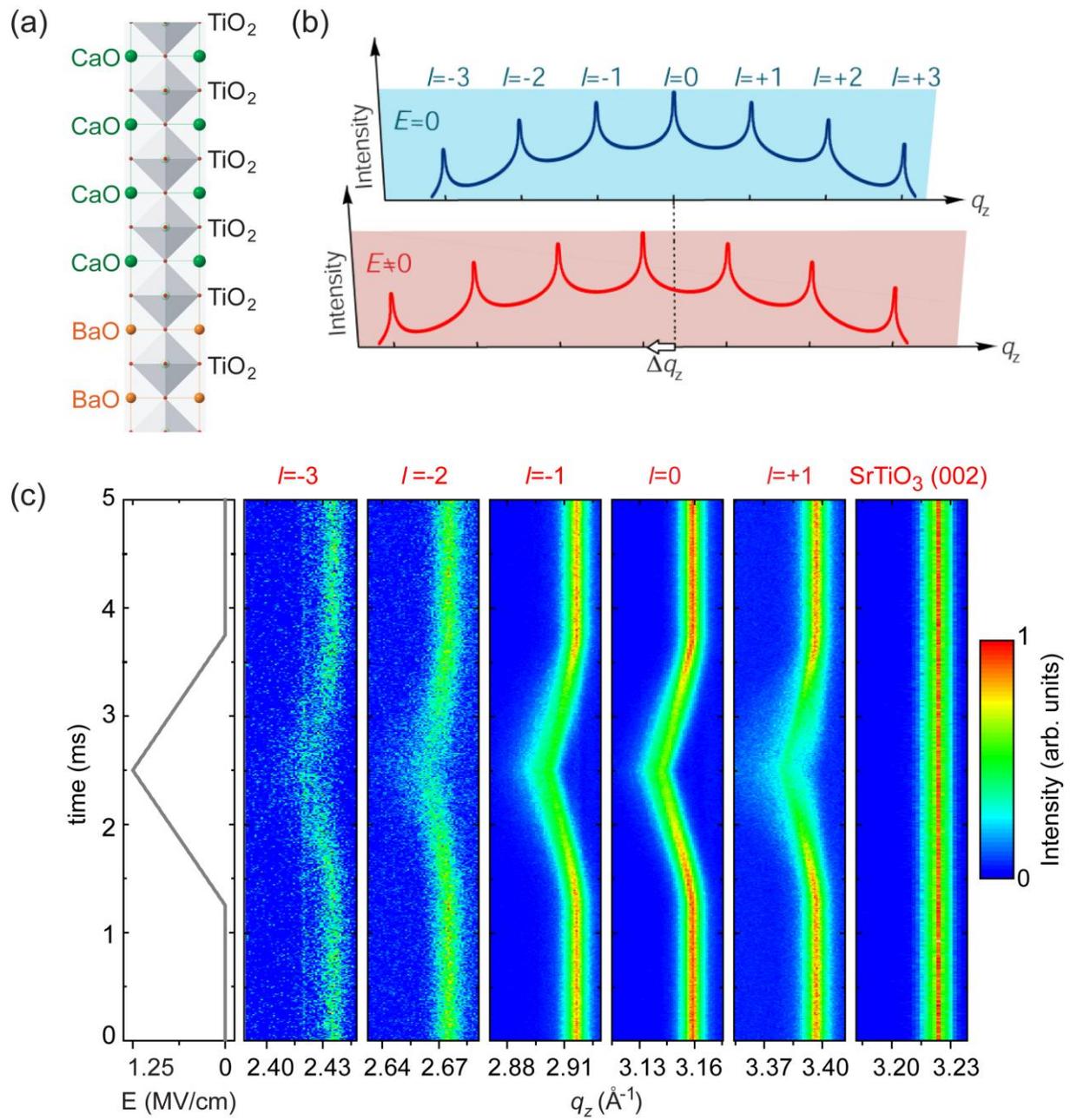

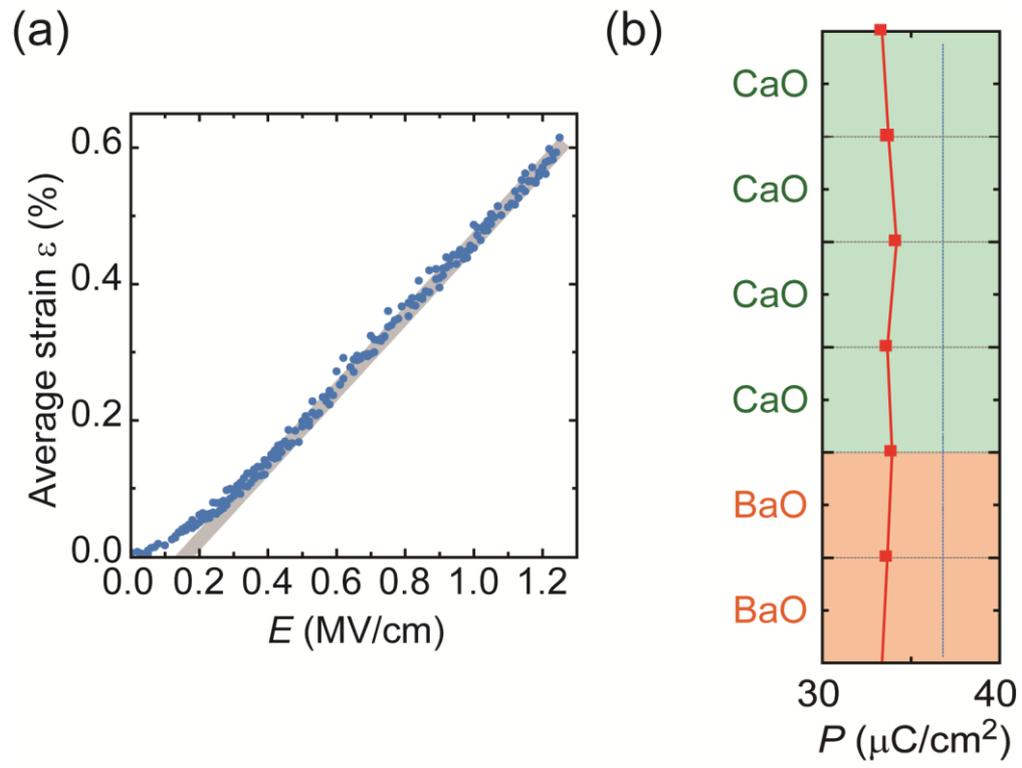

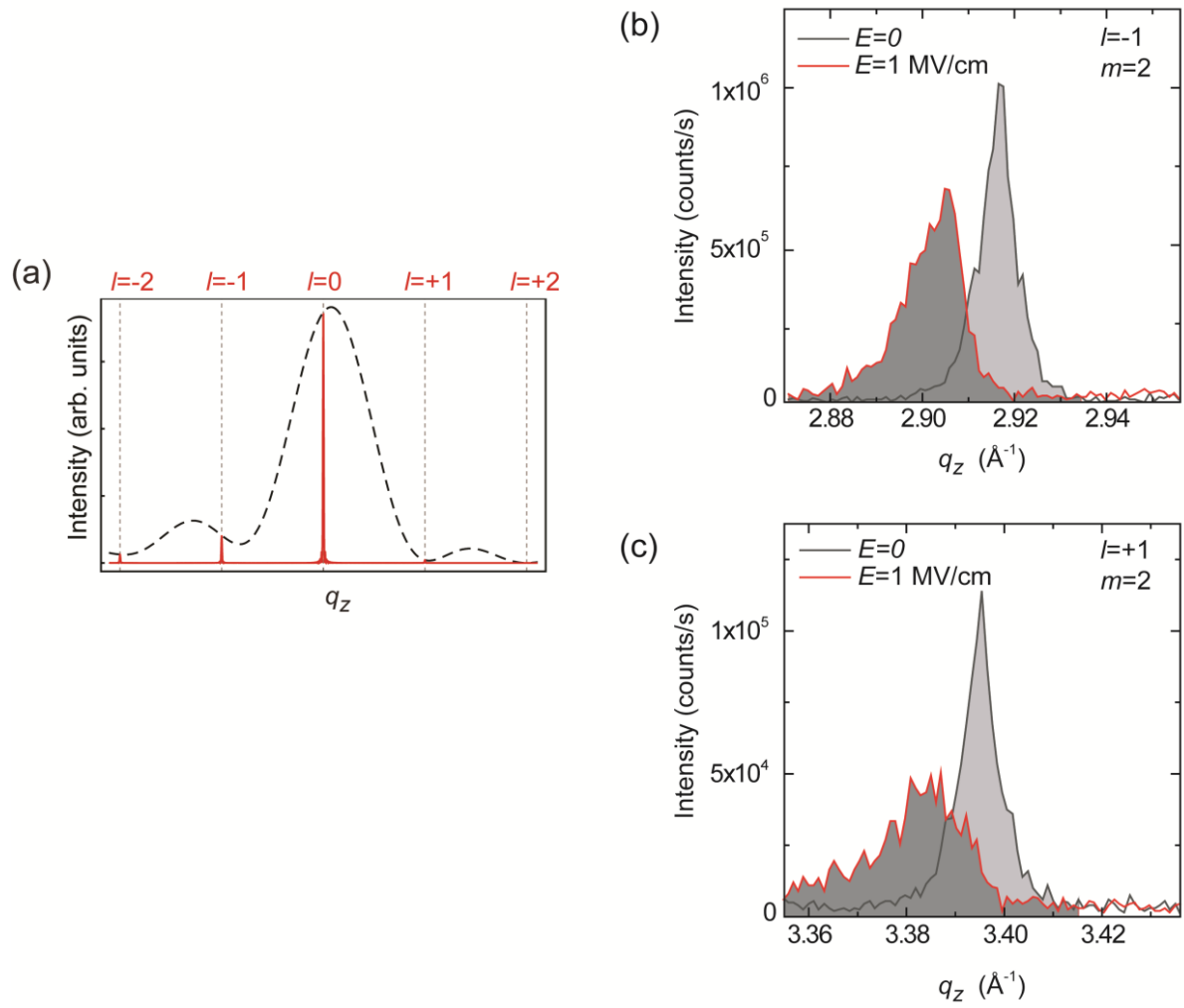

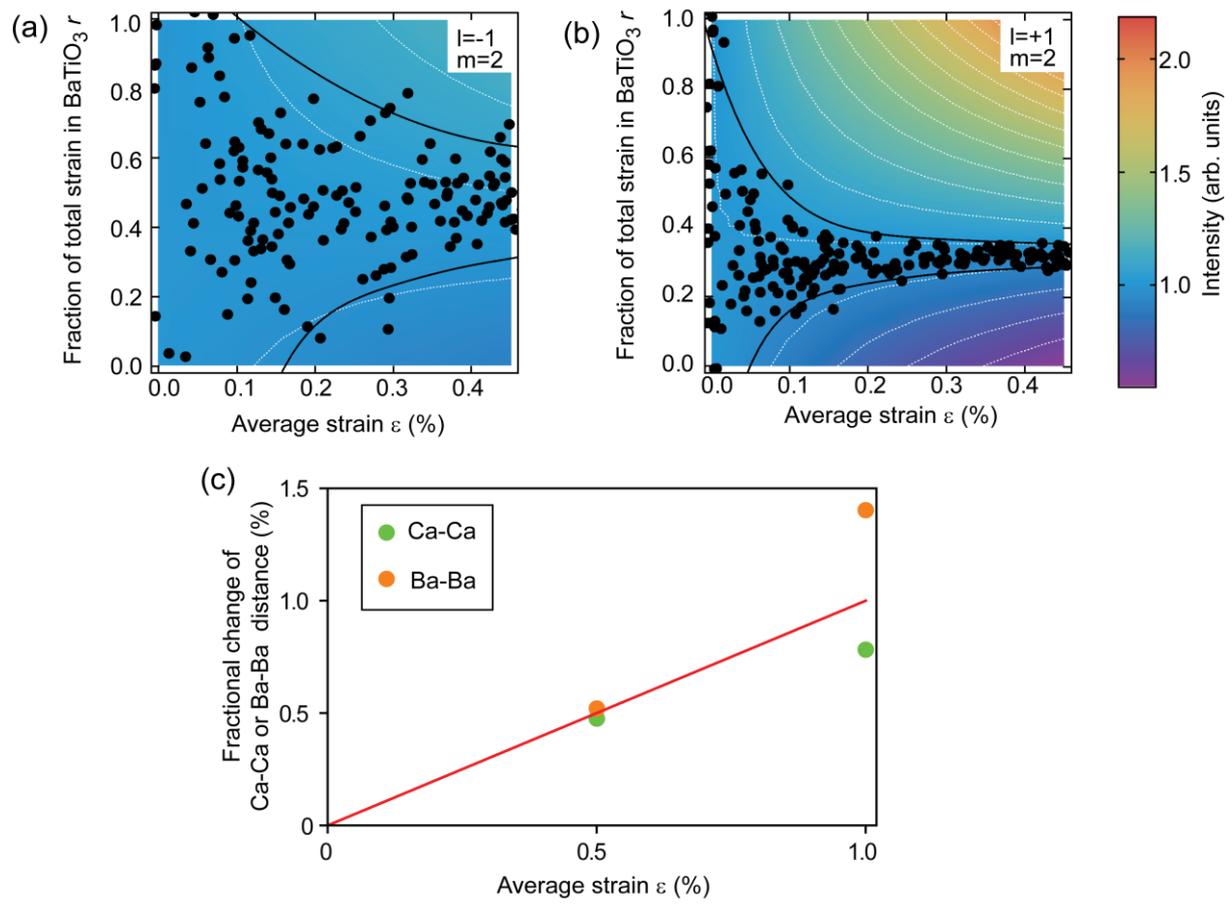